\documentclass[10pt,twocolumn,floatfix,aps,pra,superscriptaddress,nofootinbib]{revtex4-2}

\usepackage{amsmath,amssymb,amsthm,mathrsfs,amsfonts,dsfont,amstext}
\usepackage{textcomp,pbox}
\usepackage[export]{adjustbox}
\usepackage{soul} 
\usepackage{bm,xspace}
\usepackage{dcolumn,booktabs,url}
\usepackage[scaled]{helvet}
\usepackage{sansmath,gensymb}
\usepackage{tikz,graphicx,transparent,color}
\usepackage{multirow}
\usepackage[separate-uncertainty = true]{siunitx}
\usepackage{comment}
\usepackage{physics}
\usepackage{pdfpages}
\usepackage[toc,page]{appendix}

\usepackage[dvipsnames,table,xcdraw]{xcolor} 
\usepackage{array}
\usepackage{makecell}
\usepackage{tabularx}
\usepackage{booktabs}
\usepackage{pdflscape}
\usepackage{float}
\usepackage{graphicx}
\usepackage{multirow}
\usepackage{url}

\newcolumntype{L}{>{\raggedright\arraybackslash}X}
\newcolumntype{C}[1]{>{\centering\arraybackslash}m{#1}}

\makeatletter
\AtBeginDocument{\let\LS@rot\@undefined}
\def\maketitle{\@author@finish\title@column\titleblock@produce\suppressfloats[t]}
\makeatother

\usepackage{natbib} 

\usepackage[colorlinks=true]{hyperref}

\hypersetup{
     colorlinks   = true,
     citecolor    = red,
     linkcolor    = blue,
     urlcolor     = red}

\graphicspath{{./Figures/}}

\begin{document}

\title{The dynamic 4.8.8 Floquet code} 
\author{Aliki A. Capatos}
\email{aliki.anna.capatos@nbi.ku.dk}
\affiliation{Quantum Engineering Centre for Doctoral Training, University of Bristol, Bristol, United Kingdom}
\affiliation{NNF Quantum Computing Programme, Niels Bohr Institute, University of Copenhagen, Blegdamsvej 17, DK-2100 Copenhagen Ø, Denmark}

\begin{abstract}
Fault-tolerant quantum memories depend on the syndrome extraction circuit as much as on the underlying code. \emph{Ancilla-free} or \emph{dynamic} circuits are an effective way to improve this circuit layer. For the 6.6.6 honeycomb Floquet code, making the circuit dynamic raises the threshold and lowers the qubit overhead, but at the cost of halving the spatial code distance. A dynamic construction for the 4.8.8 lattice layout was conjectured to preserve full distance. I confirm this and give a dynamic measurement circuit for the CSS 4.8.8 Floquet code. To benchmark it, I construct and compare four circuit-level implementations on a torus, including two dynamic variants (with and without mid-circuit resets), the standard ancilla-based circuit, and a pipelined ancilla-based circuit. Under circuit-level depolarising noise, the reset dynamic circuit reaches a per-round threshold of $0.463\%$ $(0.490\%)$ with MWPM (BP+matching), while the no-reset variant reaches the highest threshold of all four circuits at $0.512\%$ $(0.574\%)$. The standard ancilla-based circuit only achieves $0.228\%$ $(0.240\%)$, but the pipelined schedule reaches $0.478\%$ $(0.489\%)$. The reset dynamic circuit also has a faster-growing timelike distance, with $2\le d_t/n_{\mathrm{qec}}\le 3$ asymptotically against a tight $3/2$ for the other three, and running it for fewer rounds gives the smallest spacetime volume in the fast-reset regime, while the no-reset variant is smallest in the slow-reset regime. The 4.8.8 dynamic circuits therefore see the expected threshold gain and overhead reduction without the spatial-distance cost, demonstrating the advantage of dynamic syndrome extraction in Floquet codes.
\end{abstract}

\maketitle

\section{Introduction}
\label{sec:intro}
Quantum error correction is often discussed at the level of the code, centred on the number of physical qubits, the distance, the logical operators, and the set of stabilisers or checks that define the protected subspace. In a real fault-tolerant architecture, however, the syndrome extraction circuit determines how the checks are measured, how errors propagate through the hardware, how many ancillas and time steps are required, and what detector error model the decoder sees. Consequently, two circuits for the same abstract code can have different thresholds, spacetime overheads, and effective distances. Optimising the measurement circuit is therefore an essential part of optimising the fault-tolerant memory itself.

Floquet codes are an example of how the measurement schedule can form part of what actually defines the code. They encode logical qubits using a specific time-periodic schedule of two-qubit Pauli measurements~\cite{hastinghaah2021}, where the stabilisers protecting the logical information are no longer static objects measured repeatedly, but are generated and removed as the schedule advances. Two well-known examples are the 6.6.6 honeycomb Floquet code~\cite{hastinghaah2021, gidney_honeycomb_2021} and the 4.8.8 square-octagon Floquet code~\cite{paetznick2022, davydova2022}. The two-qubit check structure makes Floquet codes work especially well with another method of optimising measurement circuits -- the use of \emph{dynamic} syndrome extraction, also referred to as \emph{ancilla-free} or \emph{morphing} circuits~\cite{shaw_optimising_2026, mcewen_relaxing_2023, eickbusch_dynamic_2024, gidney_jones_color_2023}. In such a circuit, each check is measured without a conventional dedicated ancilla~\cite{topmem, austinfowler2012} by temporarily contracting the check onto one of the data qubits in its support, measuring that qubit, and then mapping back to the original frame. For the honeycomb code, this circuit-level approach has already led to a useful improvement. A dynamic measurement circuit was introduced in which each two-qubit check is measured by a four-step \(CX-M-R-CX\) gadget acting only on the two data qubits of that check~\cite{claes2025}. This removes the ancillas, giving a \(\sim 2.5\times\) qubit saving and increasing the threshold from \(0.22\%\) to \(0.29\%\) with Minimum-Weight Perfect-Matching (MWPM) under standard depolarising noise. The catch is that, on the honeycomb lattice, the dynamic circuit halves the circuit-level spatial distance. During the gadget, stabilisers deform, and some supports contract while neighbouring supports expand. These enlarged stabilisers can touch next-nearest neighbours such that a single error can flip a pair of stabilisers that would normally be separated by two lattice steps. On the honeycomb lattice, these pairs lie along the natural logical operator support, allowing one error to advance an error chain by two steps along a logical operator. In the conclusion of~\cite{claes2025}, it is noted that the 4.8.8 lattice should not have this problem, because its same-colour check chains do not coincide with the natural logical supports. Therefore, the dynamic circuit on 4.8.8 should preserve the circuit-level spatial distance while still allowing the circuit-level performance gains of the dynamic syndrome extraction. I verify this for the periodic CSS version of the code~\cite{davydova2022, moylett2025}, and compare the threshold and spacetime volume with ancilla-based implementations.

\subsection{Reader summary}
A quick summary is provided below for the already familiar reader. The sections of this letter that follow provide more of an introduction: Section~\ref{sec:dyn} gives a brief review of dynamic circuits, and Section~\ref{sec:code} defines the 4.8.8 Floquet code. Section~\ref{sec:construction} then details its ancilla-free dynamic construction, followed by Section~\ref{sec:ancilla} describing the ancilla-based variants and the pipelined schedule. Section~\ref{sec:results} presents the distance, threshold, and spacetime-volume results, and Section~\ref{sec:conclusion} concludes. Sub-threshold results are added in the appendices, alongside details on methods and checks. Data and circuits are linked \href{https://github.com/aacpt/dynamic488}{here}.

\subsubsection*{Constructions}
I build four circuit-level implementations of the 4.8.8 Floquet code on a torus, all using the same CSS schedule~\cite{davydova2022} and the same standard depolarising noise model, keeping to nearest-neighbour-only measurements:
\begin{itemize}
      \item \textbf{Ancilla-based.} The standard baseline~\cite{davydova2022}. Each edge check has its own ancilla, prepared in $\ket{+}$ for an $XX$ check or $\ket{0}$ for a $ZZ$ check, coupled to the two neighbouring data qubits, and measured in the matching basis. The data qubits are not measured directly. 
      \item \textbf{Ancilla-based, pipelined.} Same check structure, but with the reset, gate, and measurement stages of consecutive sub-rounds overlapped to shorten the Floquet period, following the SD6 honeycomb pipelining method~\cite{gidney_honeycomb_2021}.
      \item \textbf{Dynamic, reset.} Ancilla-free, adapted from the dynamic honeycomb circuit~\cite{claes2025}. Each two-qubit check is measured by contracting it onto one of its data qubits, measuring and resetting that qubit, then undoing the contraction. The mid-circuit reset also helps with leakage~\cite{shaw_optimising_2026} and timelike protection.
      \item \textbf{Dynamic, no-reset.} The same ancilla-free idea, but with the reset removed~\cite{moylett2025}. The measured data qubit is left in its projected state before the closing $CX$.
\end{itemize}

\subsubsection*{Results}
\begin{itemize}
    \item All four variants preserve the circuit-level spatial distance as $d_{\mathrm{spatial}}=L$ (Fig.~\ref{fig:distance}(a)).
    \item Per-round thresholds under standard depolarising noise, with MWPM (BP+matching): the reset dynamic circuit reaches $0.463\%$ ($0.490\%$), the no-reset $0.512\%$ ($0.574\%$), the ancilla-based $0.228\%$ ($0.240\%$), and the pipelined ancilla-based $0.478\%$ ($0.489\%$) (Fig.~\ref{fig:threshold}).
    \item The reset dynamic circuit's timelike distance grows faster per Floquet period than that of the other variants (Fig.~\ref{fig:distance}(b)). For \(n_{\mathrm{qec}}\) noisy Floquet periods, its finite-length bounds are \begin{equation}d_{\mathrm{hyper}}=2n_{\mathrm{qec}}-3, \qquad d_{\mathrm{graph}}=3n_{\mathrm{qec}}-5,\end{equation} giving the asymptotic bracket $2\le d_t/n_{\mathrm{qec}}\le 3$. The no-reset and both ancilla-based variants have a tight asymptotic $d_t/n_{\mathrm{qec}}=3/2$ (Appendix~\ref{app:timelike-c}).
    \item At hardware-realistic gate times (Table~\ref{tab:spacetime_volume}), running the reset dynamic circuit for $n_{\mathrm{qec}}=3L/4$ Floquet periods -- to match the other variants' timelike distance -- gives the smallest spacetime volume in the fast-reset regime ($0.30\times$ the un-pipelined ancilla-based reference), followed by the no-reset variant ($0.362\times$). The no-reset variant takes the lead in the slow-reset regime ($0.245\times$).    
\end{itemize}

I use the CSS schedule rather than a schedule with $YY$ measurements because the aim is not to optimise over all possible Floquet schedules, but to test whether the dynamic circuit construction preserves spatial distance on the square-octagon geometry, and to compare the resulting dynamic circuit with the ancilla-based route. The logical support, detector structure, and $X/Z$ symmetry are easier to track, and avoiding $YY$ measurements removes the extra basis-change layers, thereby allowing for more compact pipelining.

\section{Dynamic circuits}
\label{sec:dyn}

\begin{figure*}[ht!]
\centering
\includegraphics[width=\linewidth]{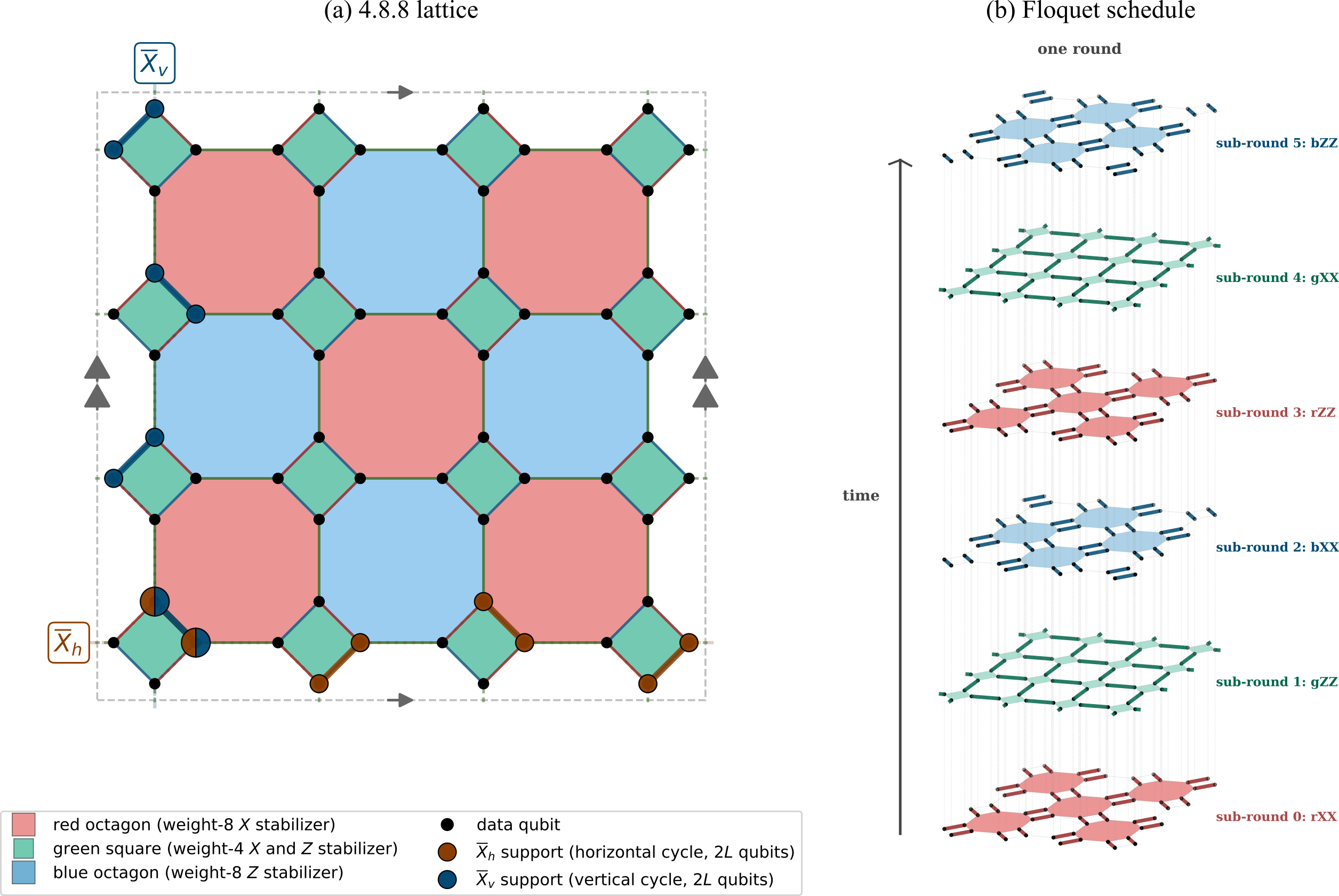}
\caption{\textbf{(a):} Part of the static 4.8.8 lattice on a torus. The data qubits (black dots) sit at the vertices of a $\{4, 8, 8\}$ tiling of the torus. Faces are 3-coloured -- red and blue octagons with green squares filling the gaps. Red/blue octagons support weight-8 plaquette stabilisers and green squares support weight-4 plaquette stabilisers. In the Floquet version of the code, the $X$/$Z$ flavour is set dynamically by the schedule. The two non-contractible cycles $\overline{X}_h$ (horizontal) and $\overline{X}_v$ (vertical) are the two independent $X$-type logical operators. The corresponding $Z$-type logicals have the same support by CSS symmetry, giving $k=2$ logical qubits encoded in $n=4L^2$ data qubits. The logicals shown have support length $2L$, while the implemented circuit-level spatial distance is $d_{\mathrm{spatial}}=L$.
\textbf{(b):} The dynamic Floquet schedule for the toric 4.8.8 code, shown as a stack of six lattice slices, one per sub-round of a single Floquet period. In each slice, the active checks (the two-qubit Pauli measurements performed during that sub-round) are shown dark in the slice's colour, with the plaquettes they infer are shaded lightly -- a green-square stabiliser at sub-round~1 ($gZZ$), a blue-octagon stabiliser at sub-round~2 ($bXX$), and so on. The schedule rotates through each face colour twice per period, once with $XX$ checks and once with $ZZ$ checks, producing a period of six sub-rounds.}
\label{fig:lattice_and_schedule}
\end{figure*}

The term \emph{Floquet} refers to the periodic check-measurement schedule that defines the code. By \emph{dynamic circuit}, I mean the ancilla-free measurement gadget used to implement one of those checks. Floquet codes are also dynamical in time, so there is some unavoidable terminology overlap, but here I keep the two ideas distinct. The various and slightly abused names (\emph{morphing}, \emph{middle-out}, \emph{dynamic}, and \emph{ancilla-free}) for these circuits all describe the same trick, whereby a check is measured via temporarily changing the stabiliser frame so that it contracts onto one of the qubits in its support, which is then measured in place of a dedicated ancilla. Closely related to the code-morphing idea first used to generate new codes~\cite{Vasmer_2022}, such circuits have already been used for several code families, for instance the surface code~\cite{mcewen_relaxing_2023, eickbusch_dynamic_2024}, the colour code~\cite{gidney_jones_color_2023}, the Bacon-Shor code~\cite{Alam2025}, bivariate bicycle codes where they lowered connectivity requirements~\cite{shaw_morphing_2025}, and the honeycomb Floquet code~\cite{claes2025, Benito_2025}, which is the direct inspiration for this work. Dynamic circuits have also been developed into a broader framework for optimising codes and syndrome extraction circuits together~\cite{shaw_optimising_2026}. 

Recall that a Clifford circuit maps stabilisers to stabilisers: \begin{equation}S_i \mapsto U S_i U^\dagger.\end{equation} If the circuit is chosen well, some part of the code can become locally fixed in the new frame and stop carrying logical information. Those qubits can then be measured or reset without measuring the logical state. A dynamic measurement circuit uses this idea only temporarily, such that the circuit changes frame, measures the now-simple check, and then changes back. For example, consider an $XX$ check on an edge between the measured endpoint $c$ and its $X$-flavour partner $a^X$. A CNOT with control $c$ and target $a^X$ gives
\begin{equation}
\mathrm{CX}_{c,a^X}\left(X_c X_{a^X}\right)\mathrm{CX}_{c,a^X}^{\dagger}=X_c .
\end{equation} 
After the first CNOT, the two-qubit $XX$ check has become a single-qubit $X_c$ check. I can then measure $c$ in the $X$ basis, reset it to $|+\rangle$, and apply the same CNOT again to return to the original data frame. This gives the 4-TICK gadget 
\begin{equation}
\mathrm{CX}\;-\;M_X\;-\;R_X\;-\;\mathrm{CX}.
\end{equation} 
The $ZZ$ gadget is the same idea, but with the CNOT direction and measurement basis chosen so that the $ZZ$ check is mapped to a single-qubit $Z$ measurement. 

\begin{figure*}[ht!]
\centering
\includegraphics[width=\linewidth]{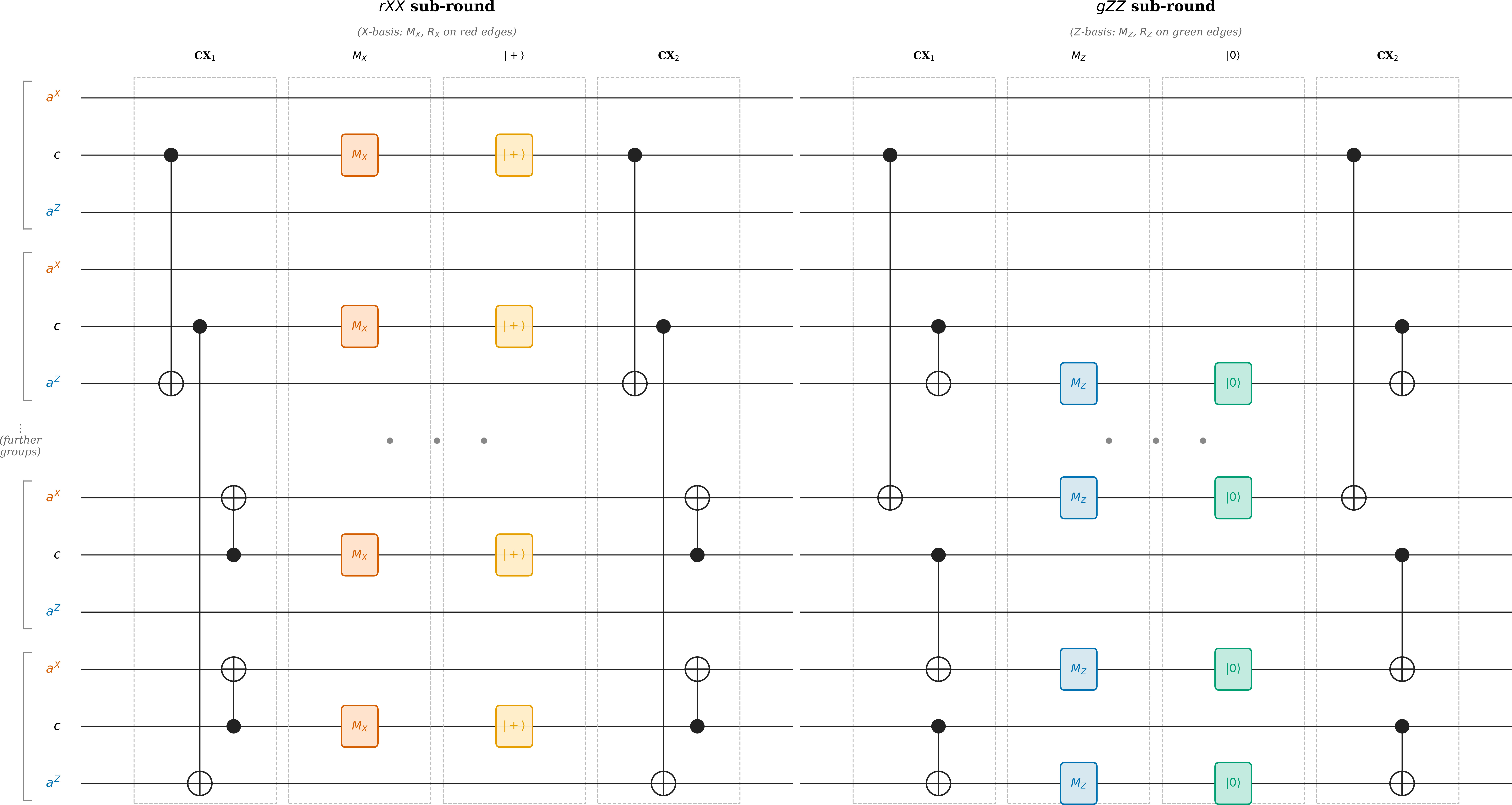}
\caption{Two consecutive sub-rounds of the reset dynamic 4.8.8 Floquet circuit. Both sub-rounds use the same 4-TICK \(CX-M-R-CX\) gadget, differing only in the basis. The left half ($rXX$) measures $XX$ checks on every red edge, in the $X$ basis. The right half ($gZZ$) measures $ZZ$ checks on every green edge, in the $Z$ basis. Because the data graph is bipartite and the two-colouring alternates which class is contracted, the measured (contracted) qubit changes between sub-rounds. In the $rXX$ half, the central qubit $c$ is measured with its red-edge neighbour $a^X$ as partner, while in the $gZZ$ half, the green-edge neighbour $a^Z$ is measured with $c$ now acting as \emph{its} partner. Each three-wire group thus shows one qubit alongside its $X$- and $Z$-flavour partners, but the groups are not disjoint -- every partner is itself the measured qubit of another group in a neighbouring sub-round, so the $CX$ gates cross group boundaries.}
\label{fig:subround-pair-circuit}
\end{figure*}

There are several reasons to use dynamic circuits. First, they can, of course, reduce the qubit count. For the 4.8.8 code, an ancilla-based circuit uses one ancilla per measured edge, yielding $6L^2$ ancillas in addition to $4L^2$ data qubits, for a total of $10L^2$. The dynamic circuit uses only the $4L^2$ data qubits, a $2.5\times$ reduction in qubit count. They can also make connectivity requirements more flexible. In many stabiliser measurement circuits, an ancilla has to couple to every qubit in the stabiliser support, but dynamic circuits can relax this constraint; the same flexibility can help with defective qubits or broken couplers~\cite{Debroy2025luciinsurfacecode, anker2025, acid, zhou2025louvrerelaxinghardwarerequirements, higgott2025handlingfabricationdefectshexgrid} because the measured qubit is chosen from inside the check support. Dynamic circuits are also a good fit for Floquet codes, in which the dynamic gadget implements the two-qubit Pauli checks on lattice edges without ancillas. The honeycomb code dynamic circuit~\cite{claes2025} can be adapted naturally to the CSS 4.8.8 schedule, as I show in Section~\ref{sec:construction}.

\section{The CSS 4.8.8 Floquet code}
\label{sec:code}
The 4.8.8 lattice is the square-octagon tiling of the plane such that at every vertex, three edges meet, and the faces alternate between squares and octagons. The faces are 3-colourable, labelled the usual red, green, and blue. Each edge inherits the colour of the same-coloured faces it joins at its two ends. On a torus of $L \times L$ unit cells, the data graph has $4L^2$ qubits and $6L^2$ edges, partitioned $2L^2$ per colour. There are two non-contractible logical $X$ cycles and two non-contractible logical $Z$ cycles, so the code encodes two logical qubits. Figure~\ref{fig:lattice_and_schedule}(a) shows a patch of the lattice.

\subsection{Checks, schedule, and ISGs}
\label{sec:terminology}
The CSS 4.8.8 Floquet code is built from a periodic sequence of two-qubit Pauli measurements, which following~\cite{davydova2022, moylett2025} I call \emph{checks}~\footnote{I have avoided using ``gauge operator" or ``gauge group" for the 4.8.8 checks simply to avoid suggesting a parent subsystem code structure~\cite{davydova2022}, as is present in the Floquet honeycomb code.}. The six checks in the period $rXX$, $gZZ$, $bXX$, $rZZ$, $gXX$, $bZZ$ do not all commute and cannot all be measured at once, so the protocol measures one commuting layer per sub-round, with all checks of a single colour in either $X$ or $Z$:
\[
\begin{array}{@{\quad}l@{\quad}l@{\quad}l@{\quad}}
\text{sub-round 0: rXX} & \text{1: gZZ} & \text{2: bXX} \\
\text{\phantom{sub-round }}\text{3: rZZ} & \text{4: gXX} & \text{5: bZZ}
\end{array}
\]
At each sub-round $r$ the prepared state is stabilised by an Abelian group $\Sigma(r)$, the \emph{instantaneous stabiliser group} (ISG), generated by the active checks just measured together with the local plaquette stabilisers -- weight-8 operators around the octagons and weight-4 operators around the green squares, $X$- or $Z$-flavoured depending on the sub-round -- whose signs are fixed by multiplying recent check outcomes. The active checks turn over every sub-round, each a stabiliser only in the sub-round it is measured; the plaquette stabilisers, being products of checks, persist for several sub-rounds until an anti-commuting check destroys them~\cite{davydova2023automorphism}. They are not elements of the static centre $Z(\langle\text{all checks}\rangle) = \langle\prod_i X_i, \prod_i Z_i\rangle$ (just the two global $\mathbb{Z}_2$ symmetries), but are dynamical objects, emerging and disappearing as the schedule advances. Repeating this period-6 cycle generates the dynamic Floquet code (Fig.~\ref{fig:lattice_and_schedule}(b)).

On the honeycomb lattice, the logical operator can be expressed as a chain of same-colour checks along a non-contractible cycle~\cite{davydova2023automorphism}. In the dynamic circuit, the contraction step causes neighbouring stabilisers to grow and come to touch their next-nearest neighbours, such that a single error can then flip pairs of stabilisers that would normally sit two lattice steps apart, and because these next-nearest neighbour pairs sit along the same-colour-check chain that supports the logical, each such error advances the error chain by two qubits along that logical. The result is that the spatial distance is halved~\cite{claes2025}. In the 4.8.8 code, this next-nearest-neighbour error mechanism does not align with the logical supports. And so, the dynamic 4.8.8 circuit should preserve the full code distance. This is verified in Section~\ref{sec:results-distance} (Figure~\ref{fig:distance}).

\section{Construction}
\label{sec:construction}
This section describes the dynamic circuit construction in three steps. First, two-colouring the lattice fixes which qubit is measured on each check edge. Second, the sub-round schedule fixes the order in which checks are measured and the gadget applied to each one. Finally, the detector group fixes which combinations of measurement outcomes are deterministic. Making sure the right detectors were being used was the hard part, done via a combination of Stim's~\cite{gidney_stim} \texttt{flow\_generators()} and a search to throw out any non-local or oversized detectors so that the remaining detector basis was local and as small as possible. The method can be readily applied to any stabiliser code.

\subsection{Two-colouring}
\label{sec:two-colouring}
The dynamic gadget needs us to pick, for every check edge, one of its two endpoints as the measured qubit each time that edge is measured. I do this by 2-colouring the data graph, alternating which colour class is measured. The 4.8.8 data graph is bipartite, with vertices split into two classes, $A$ and $B$, and every check edge connecting one $A$-qubit to one $B$-qubit. I use the colouring
\begin{equation}
  \mathrm{class}(i,j,d) = (i+j+d)\bmod 2 \in \{A,B\},
\label{eq:two-colouring}
\end{equation}
where $(i,j)$ is the unit-cell index and $d \in \{0,1,2,3\}$ labels the qubit position within the unit cell. $A$-qubits are measured during the three $XX$ sub-rounds, while $B$-qubits are measured during the three $ZZ$ sub-rounds. Because each qubit touches one edge of each colour, each data qubit is measured once for each relevant edge colour in a Floquet period. That is, in the reset variant, each data qubit is measured and reset three times per period, and every qubit is freshly initialised within any pair of consecutive sub-rounds. For each flavour $X$ or $Z$, either measurement class can be chosen, and these two binary choices give us four variants to choose from. These are equivalent, and give the same threshold (Appendix~\ref{app:gadget-mapping}).

\subsection{Sub-round schedule and gadget}
\label{sec:schedule}
Each check measurement is implemented by the 4-TICK gadget from~\cite{claes2025} on the two qubits of the check edge, with no ancilla. For $XX$, where $c$ is the measured endpoint and $a^X$ is its $X$-flavour partner, the gadget is
\[
\begin{array}{@{\quad}c@{\quad}c@{\quad}c@{\quad}c@{\quad}}
\text{TICK 1} & \text{TICK 2} & \text{TICK 3} & \text{TICK 4} \\
\text{------} & \text{------} & \text{------} & \text{------} \\
\mathrm{CX}(c,a^X) & M_X(c) & R_X(c) & \mathrm{CX}(c,a^X)
\end{array}
\]
For $ZZ$, where $a^Z$ is the $Z$-flavour partner, the $CX$ direction is reversed and $\{MX, RX\}$ replaced with $\{M, R\}$. The first $CX$ maps the check to a single-qubit Pauli on $c$. The single-qubit measurement reads its eigenvalue, and the reset prepares the post-measurement frame. The ending $CX$ maps the measured single-qubit Pauli back onto the two-qubit edge check operator, so that the recorded measurement outcome is the eigenvalue of the intended \(XX\) or \(ZZ\) check. Figure~\ref{fig:subround-pair-circuit} shows two consecutive sub-rounds of this construction. One Floquet period is 24 TICKs, $4 \text{ TICKs per sub-round} \times 6 \text{ sub-rounds}$. While the gadget is identical to that of the honeycomb, the schedule is now period 6 rather than 3, and the $A/B$ alternates across six sub-rounds rather than three.

I also build a variant without the reset step (and thus without the leakage protection), much like~\cite{moylett2025}. For \(XX\):
\[
\begin{array}{@{\quad}c@{\quad}c@{\quad}c@{\quad}}
\text{TICK 1} & \text{TICK 2} & \text{TICK 3} \\
\text{------} & \text{------} & \text{------} \\
\mathrm{CX}(c,a^X) & M_X(c) & \mathrm{CX}(c,a^X)
\end{array}
\]
Therefore, the no-reset circuit has three \(XX\) sub-rounds of 3 TICKs and three \(ZZ\) sub-rounds of 3 TICKs, putting it at 18 TICKs per Floquet period. 

\begin{figure*}[ht!]
\centering
\includegraphics[width=\linewidth]{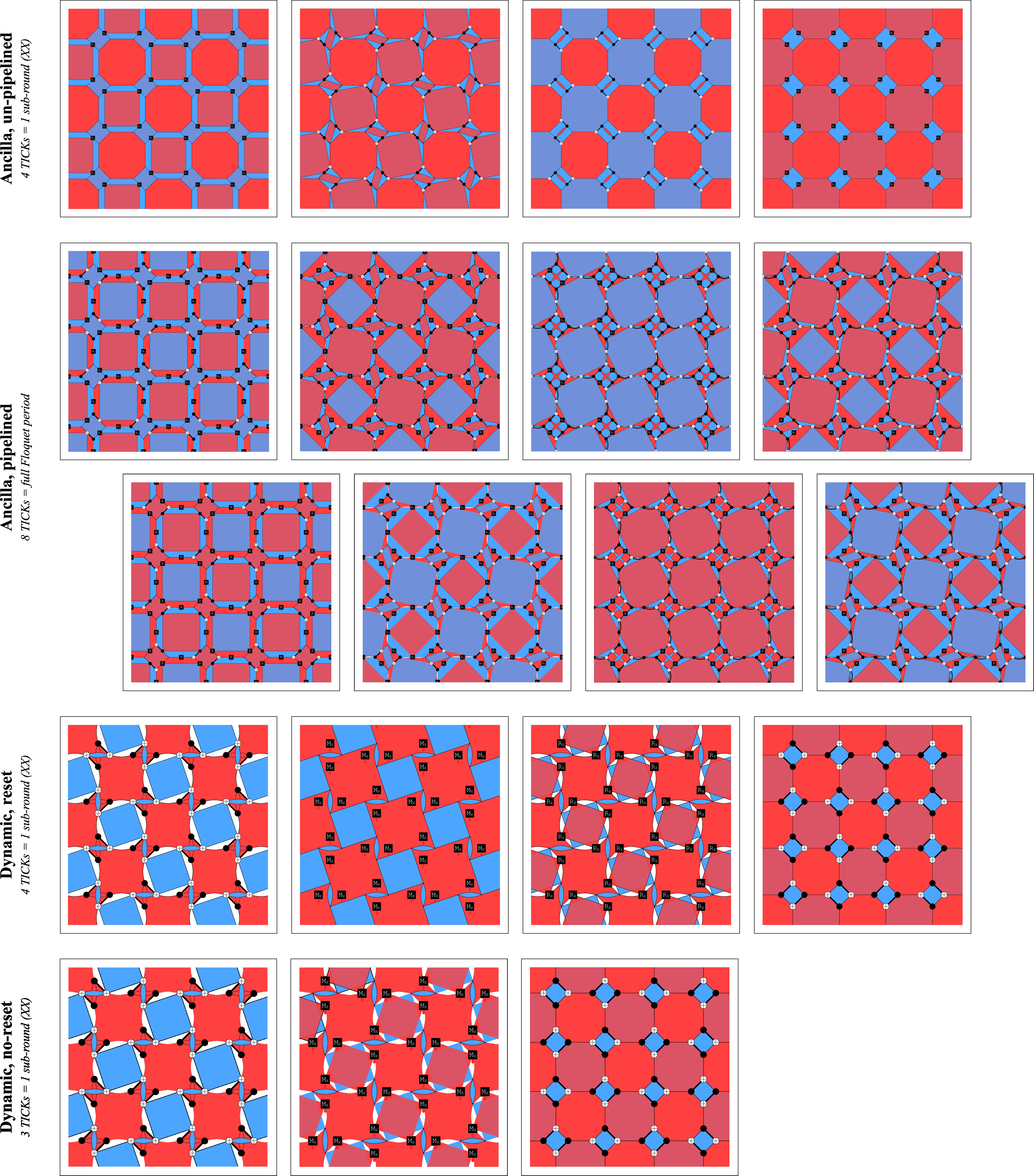}
\caption{One $rXX$ sub-round (or, for the pipelined ancilla-based circuit, one full Floquet period) of each variant's circuit of a part of the 4.8.8 lattice, showing the evolving stabiliser support. Each panel is one TICK. Stabilisers are coloured polygons, with blue for $Z$ and red for $X$. For a full period, the un-pipelined ancilla-based and reset dynamic variants go through 24 TICKs total, the no-reset dynamic 18, and the pipelined ancilla-based variant only 8. For the full period of each variant, see \href{https://github.com/aacpt/dynamic488}{here}.
\textbf{Row~1 (ancilla-based, un-pipelined):} one $XX$ check measurement in four TICKs -- $R$ on the ancillas, $\text{CX}_1$, $\text{CX}_2$, $M$ on the ancillas.
\textbf{Row~2 (ancilla-based, pipelined):} one full Floquet period in 8 TICKs, split across two rows. There is no isolated $XX$ sub-round in the pipelined schedule because three sub-rounds are in flight per TICK, and the stabiliser pattern rotates through all six sub-rounds across the window.
\textbf{Row~3 (dynamic, reset):} the $XX$ check measurements implemented without an ancilla, using the 4-TICK \(CX-M_X-R_X-CX\) gadget. As each edge check operator is contracted onto one endpoint, neighbouring stabilisers deform, grow, and return to their original supports after the closing $CX$.
\textbf{Row~4 (dynamic, no-reset):} the same $XX$ check measurements implemented without a reset between sub-rounds, using the 3-TICK \(CX-M-CX\) gadget.}
\label{fig:detectors}
\end{figure*}

\subsection{Detectors}
\label{sec:detectors}
The detectors for the two dynamic variants are found in different ways, because the reset changes which outcomes are deterministic. A detector is a set of check outcomes whose combined parity is fixed in the absence of noise. The natural detectors of a CSS Floquet code are plaquette inferences, where the same-flavour checks around a plaquette multiply to reconstruct that plaquette's stabiliser, and the $XOR$ of two such inferences of the same stabiliser is deterministic without errors. The period-six schedule reads out each plaquette stabiliser twice per period, but from different-coloured edges and at uneven spacings. Consecutive readings are separated by either four or two sub-rounds. The blue-octagon $X$ stabiliser, for example, is read at sub-round~0 ($rXX$, from its red edges) and again at sub-round~4 ($gXX$, from its green edges). Over that four-sub-round gap, every intervening check commutes with the stabiliser, so it is left undisturbed, and the $XOR$ of the two readings is a valid detector. The other gap is only two sub-rounds (sub-round~4 to sub-round~0 of the next period), and it straddles an anti-commuting check ($bZZ$) that randomises the stabiliser, so it gives no detector. Each plaquette and flavour therefore produces exactly one detector per period on its gap-four transition. For the no-reset variant, I read these gap-four detectors straight off the schedule, as~\cite{moylett2025} do for their no-reset 6.6.6 code.

This does not work for the reset variant. Resetting the measured qubit after each measurement breaks the link between one inference and the next, so the gap-four $XOR$ is no longer deterministic. Instead of attempting to find the reset variant's detectors by hand, I feed the detectors found by Stim into a \emph{greedy $XOR$-pairwise detector reducer} to find something optimal. In Stim, a detector is a stabiliser flow of the noiseless circuit whose Pauli input and output are both trivial, so its measurement support is a deterministic combination of outcome bits in the absence of noise. I get a generating set of such flows from Stim's \texttt{flow\_generators()}, keep only the trivial-in/trivial-out flows as detectors (discarding the $n$ open stabiliser flows with non-trivial output), and emit them as \texttt{DETECTOR} instructions. This ensures everything is deterministic, but not that the resulting detector basis is local or physically natural. To enforce this, some reduction is needed to keep the detectors minimum-weight, especially since the raw \texttt{flow\_generators} output has some very high weights. For instance, at $L = 4$, $n_{\mathrm{qec}} = 4$ the raw basis has mean detector weight 19.9 and maximum 80. I reduce it using greedy $XOR$-pairwise reduction, which repeatedly tries to replace $D_i$ with $D_i \oplus D_j$ if the resulting weight is lower, until no further reduction is possible. I also drop the topological detectors, those corresponding to non-contractible loops on the torus that are not useful for memory experiments (Appendix~\ref{app:nonlocal-detectors}). After reduction and dropping the topological detectors, the basis now has weights only in $\{4, 8\}$. Figure~\ref{fig:detectors} shows the first $rXX$ sub-round for each variant. Appendix~\ref{app:detector-audit} contains further details, and the full set of detslices showing each variant and its detectors over a whole Floquet period is \href{https://github.com/aacpt/dynamic488}{linked}.

\begin{figure*}[ht!]
\centering
\includegraphics[width=\linewidth]{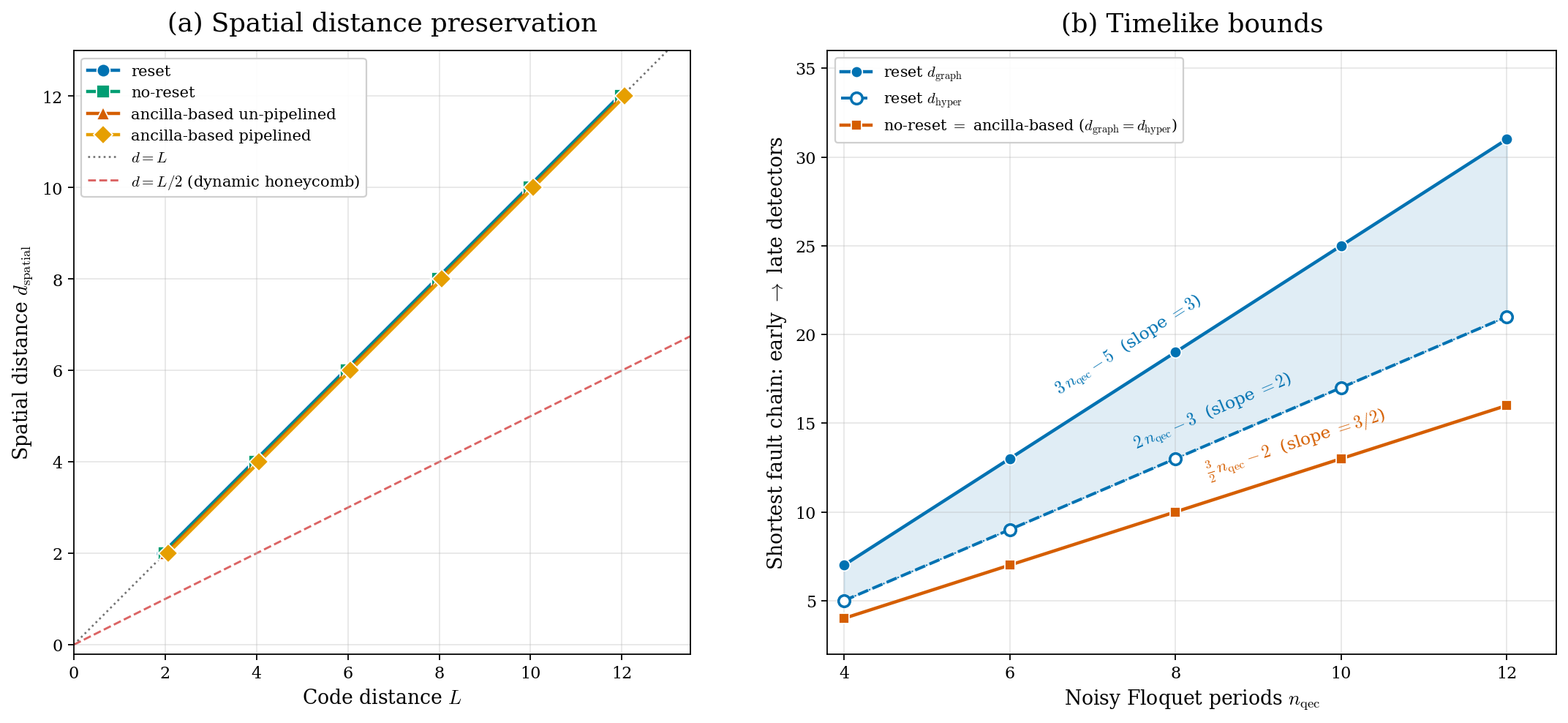}
\caption{\textbf{(a):} Spatial distance $d_{\mathrm{spatial}}$ versus $L$ for the four 4.8.8 variants. $n_{\mathrm{warmup}} = n_{\mathrm{tail}} = 2$ and $n_{\mathrm{qec}} = L$.
\textbf{(b):} Timelike-distance bounds for the four variants. The slope of this chain length versus $n_{\mathrm{qec}}$ gives $d_t / n_{\mathrm{qec}}$, the timelike distance per Floquet period~\cite{claes2025}; one 4.8.8 period contains six sub-rounds. Filled markers and solid lines show $d_{\mathrm{graph}}$, which ignores hyperedges and gives an upper bound on $d_t$. The $x$-axis starts at 4 because fewer rounds introduce boundary effects.}
\label{fig:distance}
\end{figure*}

\section{Ancilla-based pipelining}
\label{sec:ancilla}

\begin{table}[t]
\centering
\begin{tabular}{c|cccccccc}
\hline
\textbf{Gadget} & \multicolumn{8}{c}{\textbf{Compressed TICK offset}} \\
 & $0$ & $1$ & $2$ & $3$ & $4$ & $5$ & $6$ & $7$ \\
\hline
$rXX$ & $R$ & $\mathrm{CX}_a$ & $\mathrm{CX}_b$ & $M$ & $\cdot$ & $\cdot$ & $\cdot$ & $\cdot$ \\
$gZZ$ & $\cdot$ & $R$ & $\mathrm{CX}_a$ & $\mathrm{CX}_b$ & $M$ & $\cdot$ & $\cdot$ & $\cdot$ \\
$bXX$ & $\cdot$ & $\cdot$ & $R$ & $\mathrm{CX}_a$ & $\mathrm{CX}_b$ & $M$ & $\cdot$ & $\cdot$ \\
$rZZ$ & $\cdot$ & $\cdot$ & $\cdot$ & $\cdot$ & $R$ & $\mathrm{CX}_a$ & $\mathrm{CX}_b$ & $M$ \\
$gXX$ & $M$ & $\cdot$ & $\cdot$ & $\cdot$ & $\cdot$ & $R$ & $\mathrm{CX}_a$ & $\mathrm{CX}_b$ \\
$bZZ$ & $\mathrm{CX}_b$ & $M$ & $\cdot$ & $\cdot$ & $\cdot$ & $\cdot$ & $R$ & $\mathrm{CX}_a$ \\
\hline
\end{tabular}
\caption{Decomposition of the pipelined 4.8.8 measurement schedule into gadget identities. Each row corresponds to one Floquet sub-round gadget, and each column is a compressed TICK offset modulo 8. The four phases of each gadget are $R \to \mathrm{CX}_a \to \mathrm{CX}_b \to M$. Rows whose phases cross the end of the compressed period are shown modulo 8. For example, the $gXX$ gadget runs chronologically as $R_5 \to \mathrm{CX}_{a,6} \to \mathrm{CX}_{b,7} \to M_8$, so its measurement appears in the $0$ column. Similarly, $bZZ$ runs as $R_6 \to \mathrm{CX}_{a,7} \to \mathrm{CX}_{b,8} \to M_9$, which appears as $R_6 \to \mathrm{CX}_{a,7} \to \mathrm{CX}_{b,0} \to M_1$. As such, the entries in the last two rows indicate gadgets wrapping around the period boundary.}
\label{tab:pipelined-exploded}
\end{table}

While the ancilla-free dynamic implementation offers advantages over the standard ancilla-based approach, the application of \emph{pipelining}~\cite{gidney_honeycomb_2021} may compensate for this. Pipelining overlaps multiple sub-rounds in time, compressing the total number of time steps needed. In order to compare the pipelined version against a non-pipelined ancilla-based case, I implement the standard ancilla-based 4.8.8 circuit following~\cite{davydova2022}, where each check measurement uses one ancilla prepared in $|+\rangle$ for $XX$ or $|0\rangle$ for $ZZ$, then two $CX$s onto the check's data endpoints, then a single-qubit measurement of the ancilla. Each sub-round is 4 TICKs (\(R, CX, CX, M\)), giving 24 TICKs per Floquet period, the same depth as the dynamic circuit. The qubit overhead is $2.5\times$ as many physical qubits ($10L^2$ total).

The SD6 honeycomb circuit~\cite{gidney_honeycomb_2021} achieves 2 TICKs per sub-round (a $2\times$ depth reduction over the naive 4). The 4.8.8 code has two nice structural properties, the first of which allows pipelining to be applied, and the second of which means we can be a little more aggressive with our packing:
\begin{enumerate}
  \item \textbf{Bipartite data graph.} The first $CX$ of each check measurement acts only on the parity-0 endpoints of the active colour's edges; the second $CX$ only on parity-1 endpoints. Two different sub-rounds can therefore share a TICK without data-qubit conflicts.
  \item \textbf{No $YY$ measurements.} The honeycomb's $XX \to YY \to ZZ$ schedule needs a $C_{ZYX}$ basis-change rotation on data qubits between sub-rounds, adding an extra TICK per pipeline step. The CSS schedule has no $YY$, and avoids that cost.
\end{enumerate}

The result is three sub-rounds in flight per TICK, giving exactly 8 TICKs per period -- a $3\times$ depth reduction. The pipelined circuit preserves $d = L$ at all tested distances. The schedule is laid out in Table~\ref{tab:pipelined-exploded}, with further schedule details in Appendix~\ref{app:pipelining}.

\section{Results}
\label{sec:results}
\subsection{Spatial distance}
\label{sec:results-distance}
Figure~\ref{fig:distance}(a) shows $d_{\mathrm{spatial}}$ vs $L$ for each variant. $L$ is the linear size of the torus, where the code is defined on an $L \times L$ grid of square-octagon unit cells, giving $n = 4L^2$ data qubits and $k = 2$ logical qubits. The distance $d$ is the \emph{circuit-level} distance, the smallest number of circuit errors that together cause an undetected logical error. I confirm $d_{\mathrm{spatial}} = L$ for all four variants using two independent methods on the noisy circuit. First, after dropping the non-local detectors (Appendix~\ref{app:nonlocal-detectors}), Stim's \texttt{shortest\_graphlike\_error} returns the minimum-weight graphlike logical fault directly. I confirm this on both independent non-contractible $X$-cycles (horizontal and vertical). Second, since the graphlike method ignores errors that flip three or more detectors, I use Stim's \texttt{search\_for\_undetectable\_logical\_errors}, a full circuit-error search with no graphlike approximation. This gives $d = L$ for all four variants, including the dynamic. The CSS $X \leftrightarrow Z$ symmetry of the code implies the same distance for the $Z$-type logicals.

\subsection{Timelike distance}
\label{sec:results-timelike}
The timelike distance $d_t$ of a memory experiment is the smallest number of physical errors that can produce an undetected logical error spanning the full QEC region. Following previous methods~\cite{claes2025}, I estimate $d_t$ by measuring two shortest-path distances between early-time and late-time detectors in the detector error model (DEM). The upper bound $d_{\mathrm{graph}}$ is obtained from the syndrome graph after ignoring hyperedges, i.e. errors that flip $\geq 3$ detectors, while the lower bound $d_{\mathrm{hyper}}$ is obtained by expanding each hyperedge into a clique (a graph where every pair of vertices is connected by an edge), which can introduce shortcut paths not necessarily realised by physical error chains. By construction, $d_{\mathrm{hyper}} \leq d_t \leq d_{\mathrm{graph}}.$ Both bounds grow linearly with the number of noisy Floquet periods $n_{\mathrm{qec}}$, and their slopes give corresponding bounds on $d_t/n_{\mathrm{qec}}$. 

Across the four variants, the timelike-distance data are split into two clear cases (Figure~\ref{fig:distance}(b)). For the reset dynamic circuit, the shortest paths have a simple finite-length form, 
\begin{equation}
\begin{aligned}
d_{\mathrm{graph}}(n_{\mathrm{qec}}) &= 3n_{\mathrm{qec}} - 5, \\
d_{\mathrm{hyper}}(n_{\mathrm{qec}}) &= 2n_{\mathrm{qec}} - 3.
\end{aligned}
\end{equation}
Equivalently, 
\begin{equation}2 - \frac{3}{n_{\mathrm{qec}}} \leq \frac{d_t}{n_{\mathrm{qec}}} \leq 3 - \frac{5}{n_{\mathrm{qec}}},
\end{equation} 
so, asymptotically, 
\begin{equation}
  2 \le \frac{d_t}{n_{\mathrm{qec}}}
  \le 3,
\end{equation}  
which aligns with Eq.~(19) in~\cite{shaw_optimising_2026} (Appendix~\ref{app:timelike-c}). The early and late detector windows, warmup/tail periods, and detector-phase alignment contribute constant offsets.

The remaining three variants, no-reset, ancilla-based un-pipelined, and ancilla-based pipelined, collapse to a single tight bound. Their two shortest-path estimates coincide, $d_{\mathrm{graph}} = d_{\mathrm{hyper}}$, with slope exactly 
\begin{equation}
    \frac{d_t}{n_{\mathrm{qec}}} = \tfrac{3}{2}.
\end{equation}
Thus, only the reset dynamic variant shows an increased timelike-distance bound relative to the ancilla-based circuits, with a finite-window gain of $1.33$–$2\times$ per Floquet period. The reset behaviour is similar to the improvement observed when moving from the standard ancilla honeycomb code, though the two codes must be compared per sub-round rather than per period, since the 4.8.8 Floquet period is six sub-rounds, compared with the honeycomb's three. In per-sub-round terms, the ancilla-based and no-reset variants run at $d_t/sub-round=1/4$, matching the standard honeycomb circuit's $0.75/3=1/4$~\cite{claes2025}, while the reset circuit reaches $1/3 - 1/2$, comparable to the dynamic honeycomb's $1/3 - 2/3$ (its $1 \le d_t/n_{\mathrm{qec}} \le 2$ over a three-sub-round period). The reset gadget thus delivers, on the 4.8.8 lattice, the same kind of timelike-rate improvement that the dynamic circuit delivers on the honeycomb.

To see why the no-reset variant looks like the ancilla-based variants for timelike distance, see Figure~\ref{fig:shortest_path} in Appendix~\ref{app:timelike-c}. 

\begin{figure*}[ht!]
\centering
\includegraphics[width=\linewidth]{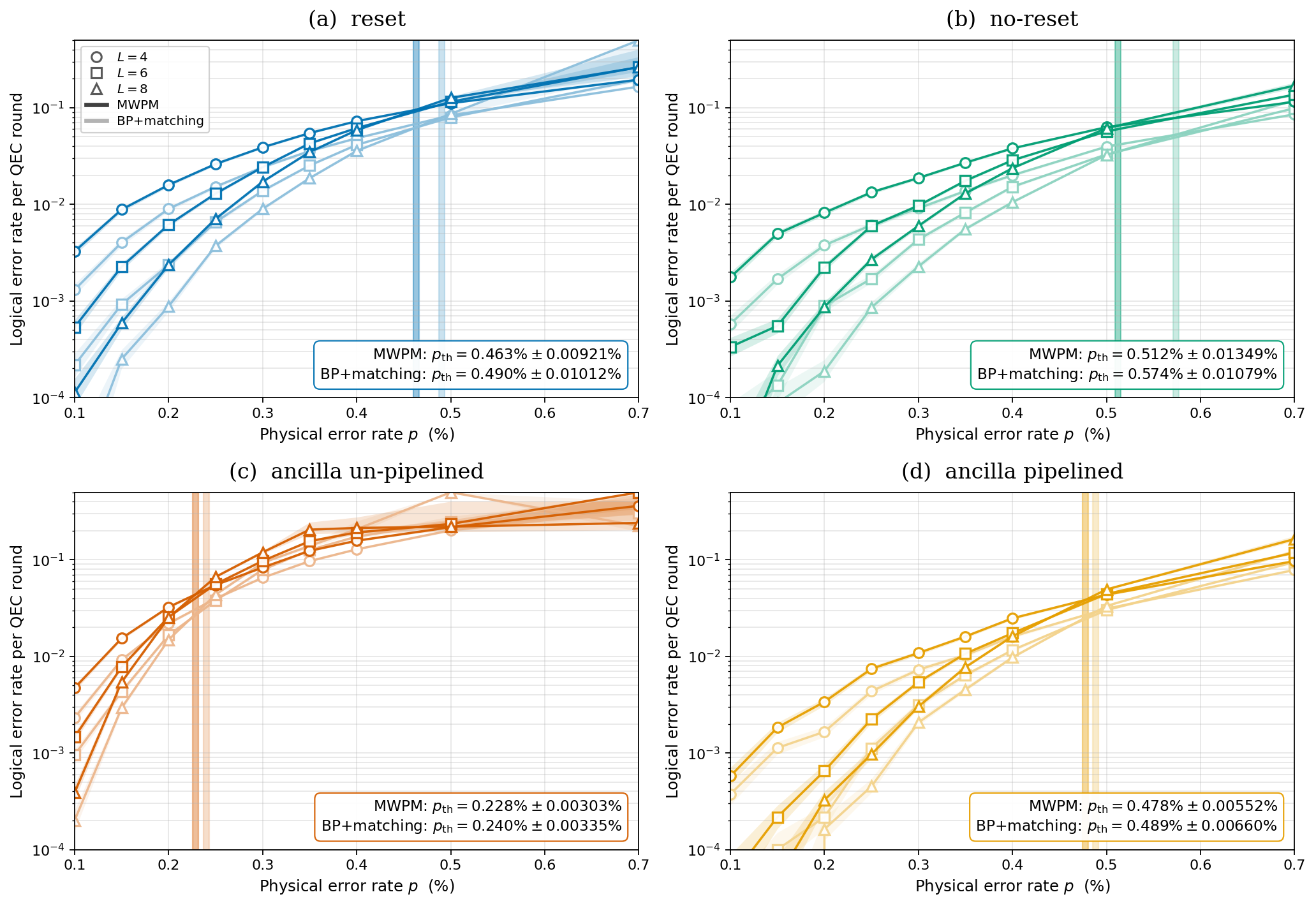}
\caption{Logical error rate per QEC round vs physical error rate $p$ for a single pair of logical operators. MWPM (dark) and BP+matching (light) are run for $10^4$ shots.}
\label{fig:threshold}
\end{figure*}

\subsection{Thresholds}
\label{sec:results-threshold}
Thresholds are estimated via the usual Monte Carlo sampling at multiple $(L, p)$ points, before decoding with PyMatching~\cite{pymatching} or BP+matching~\cite{beliefmatching}. I use $10^4$ shots per point, $n_{\mathrm{qec}} = L$ noisy periods, and two warmup periods and two tail periods. The noise model is the standard depolarising one, i.e. every one- and two-qubit gate is followed by the corresponding depolarising channel of strength $p$, idle qubits receive single-qubit depolarising noise of strength $p$, and preparation and measurement outcomes are flipped with the same probability $p$. Noise is injected only in the main region of the circuit, such that warmup and tail periods are noiseless to give clean detector boundary conditions. 

As per Figure~\ref{fig:threshold}, the reset dynamic circuit has a threshold of $p_{\mathrm{th}} = 0.463\%$ ($0.490\%$) with MWPM (BP+matching), well above the standard ancilla-based circuit at $0.228\%$ ($0.240\%$). The no-reset variant achieves the highest threshold of all four circuits, $0.512\%$ ($0.574\%$), beating both the un-pipelined ancilla baseline and the reset construction, though without the leakage protection that the latter's resets provide. The pipelined ancilla-based variant reaches $0.478\%$ ($0.489\%$) -- comparable to the reset dynamic circuit and below the no-reset variant -- but uses $2.5\times$ as many physical qubits and provides no added leakage protection. Belief propagation with matching improves on plain uncorrelated matching by using error correlations that uncorrelated matching ignores. The gain is largest for the no-reset circuit, attributed to the data qubits never being reset after they are measured, so a single fault remains across sub-rounds and connects detectors that matching treats separately.

\subsection{Spacetime volume}
\label{sec:results-spacetime}
The threshold is not the only useful metric for comparing syndrome extraction circuits. A circuit may have a higher threshold but still require more qubits or take longer to complete an error correction cycle. I compare the spacetime volume of a two-logical-qubit memory block, 
\begin{equation}V(L)=Q\,n_{\rm qec}\,t_{\rm period}, 
\end{equation} 
where \(Q\) is the number of physical qubits, \(n_{\rm qec}\) is the number of noisy Floquet periods, and \(t_{\rm period}\) is the time for one of those periods. The dynamic circuits use only the data qubits, $Q_{\rm dyn}=4L^2,$ whereas the ancilla-based circuits use one ancilla per measured edge, $Q_{\rm anc}=10L^2$. As a first comparison, suppose every TICK has the same duration. At fixed \(n_{\rm qec}=L\) and $t_{\rm period}=n_{TICKs}$,
\begin{equation}
\begin{aligned}
    V_{\rm reset}             &= 4L^2\cdot L\cdot 24 = 96L^3,\\
    V_{\rm no\text{-}reset}   &= 4L^2\cdot L\cdot 18 = 72L^3,\\
    V_{\rm anc}               &= 10L^2\cdot L\cdot 24 = 240L^3,\\
    V_{\rm anc,piped}         &= 10L^2\cdot L\cdot 8  = 80L^3.
\end{aligned}
\end{equation}
On this equal-TICK metric, the pipelined ancilla-based circuit has \(0.83\times\) the spacetime volume of the reset dynamic circuit, but only by using \(2.5\times\) as many qubits.

This fixed-\(n_{\rm qec}=L\) comparison is conservative for the reset dynamic circuit, because its timelike distance grows faster per Floquet period. When run for \(L\) periods, the no-reset and ancilla-based variants have 
\begin{equation}
  d_t(L) = \left\lfloor \tfrac{3}{2} L \right\rfloor - 2.
  \label{eq:dt-noreset}
\end{equation}
whereas the reset dynamic circuit obeys the lower bound 
\begin{equation}
d_t^{\rm reset}\geq 2n_{\rm qec}-3.
\end{equation} 
A conservative finite-size timelike-distance match then requires 
\begin{equation}
2n_{\rm qec}^{\star}-3\geq \frac{3}{2}L-2, \qquad n_{\rm qec}^{\star} = \left\lceil \frac{3L}{4}+\frac{1}{2}\right\rceil.
\end{equation} 
Asymptotically, this reduces the reset circuit's effective volume from 
\begin{equation}
V_{\rm reset}^{\rm fixed}=96L^3
\end{equation} 
down to 
\begin{equation}
V_{\rm reset}^{\rm matched} \sim 4L^2\cdot \frac{3L}{4}\cdot 24 = 72L^3.
\end{equation} 
The equal-TICK comparison with the pipelined ancilla-based circuit then flips: 
\begin{equation}
\frac{V_{\rm anc,piped}}{V_{\rm reset}^{\rm matched}} = \frac{80}{72} \approx 1.11.
\end{equation} 
Even using only the lower bound on \(d_t^{\rm reset}\), the timelike-distance-matched reset dynamic circuit has the smaller asymptotic spacetime volume.

For superconducting hardware, measurement and reset times are much slower than one- and two-qubit gates, so an 8-TICK schedule is not necessarily three times faster than a 24-TICK schedule. Table~\ref{tab:spacetime_volume} uses the superconducting-inspired operation times listed in~\cite{toreset} to illustrate something more realistic. The circuits are compiled to a \(CZ\)-native gate set, with \(\mathrm{CX}(c,t)=H_t\,\mathrm{CZ}\,H_t\). \(H\equiv S\,\sqrt{X}\,S\) up to global phase; the two \(S\) rotations are virtual-\(Z\) frame shifts that incur no physical pulse time, so each \(H\) is the duration of one physical \(\sqrt{X}\) pulse, \(20~\mathrm{ns}\) in total. Any \(H\) immediately preceding an unconditional reset is deleted because the reset erases the qubit state. In this hardware-timed model, the reset dynamic and un-pipelined ancilla-based circuits have the same per-period timing. Their ratio is therefore independent of the reset duration (\(t_{\rm period}\) cancels, leaving only the qubit-count and \(n_{\rm qec}\) factors). The same cancellation applies to the timelike-distance-matched reset curve. The no-reset and pipelined ancilla-based ratios do depend on the reset duration, because their per-period timings differ from the un-pipelined ancilla-based reference. In the slow-reset regime, the no-reset dynamic circuit has the smallest spacetime volume, about \(0.24\times\) the un-pipelined ancilla-based reference. In the fast-reset regime, the asymptotically timelike-distance-matched reset dynamic circuit has the edge, approaching \(0.30\times\). The pipelined ancilla-based circuit remains more expensive than the dynamic variants in both regimes, at \(0.77\)-\(0.83\times\) the reference.

\begin{table}[ht!]
\centering
\begin{tabular}{lcc}
\toprule
Variant & Slow reset & Fast reset \\
\midrule
ancilla-based, un-pipelined         & 1.000 & 1.000 \\
ancilla-based, pipelined            & 0.774 & 0.825 \\
reset dynamic                       & 0.400 & 0.400 \\
reset, timelike distance-matched    & 0.300 & 0.300 \\
dynamic no-reset                    & 0.245 & 0.362 \\
\bottomrule
\end{tabular}
\caption{
Spacetime volume of the 4.8.8 syndrome-extraction circuits on superconducting hardware relative to the ancilla-based un-pipelined reference (steady-state noisy volume, excluding warmup/tail). Operation durations follow the superconducting model in~\cite{toreset}: \(H=20~\mathrm{ns}\), \(\mathrm{CZ}=40~\mathrm{ns}\), \(M=600~\mathrm{ns}\), and unconditional reset is either \(R=500~\mathrm{ns}\) for slow resets or \(R=100~\mathrm{ns}\) for fast resets.}
\label{tab:spacetime_volume}
\end{table}

\section{Conclusion}
\label{sec:conclusion}
In this work, I examine dynamic circuits for the CSS 4.8.8 Floquet code. Unlike the dynamic honeycomb code, the 4.8.8 geometry preserves the full circuit-level spatial distance, so removing the ancillas gives the expected qubit savings with no trade-off. The reset variant wins on timelike-distance growth and has leakage protection, and the no-reset variant wins on threshold and spacetime volume in a slow-reset regime. Both beat the un-pipelined ancilla-based circuit. The pipelined ancilla-based schedule, however, is comparable to the reset variant on threshold but falls short of the no-reset variant, and at realistic gate times, its measurement and reset costs put its spacetime volume above both dynamic variants.

I worked entirely on a torus -- translating the dynamic 4.8.8 circuit to a planar patch with boundaries~\cite{paetznick2022, gidney_planar_2022} would facilitate stability experiments~\cite{Gidney2022stability} that would find the exact timelike distance. More broadly, Floquet code logical operations via folds and twists~\cite{moylett2025} would benefit from the increased threshold and reduced overhead reported here.

\section*{Code and circuit availability}
\label{sec:code-availability}
Data, code, etc are available at~\cite{github}, including example $L=4$ Crumble circuits of the four variants and detslices of their full Floquet periods.

\section*{Acknowledgments} 
Thank you to Jahan Claes for his advice with this project, and to João Ramos for helpful discussions. Thank you to my supervisor, Stefano Paesani, for the sanity checks and research freedom. I am grateful to the UK EPSRC (EP/S023607/1) for financial support.

\bibliography{ref.bib}

\appendix
\renewcommand{\thesubsection}{\thesection(\alph{subsection})}

\section{Sub-threshold scaling}
\label{app:subthreshold}

\begin{figure*}[ht!]
\centering
\includegraphics[width=0.92\linewidth]{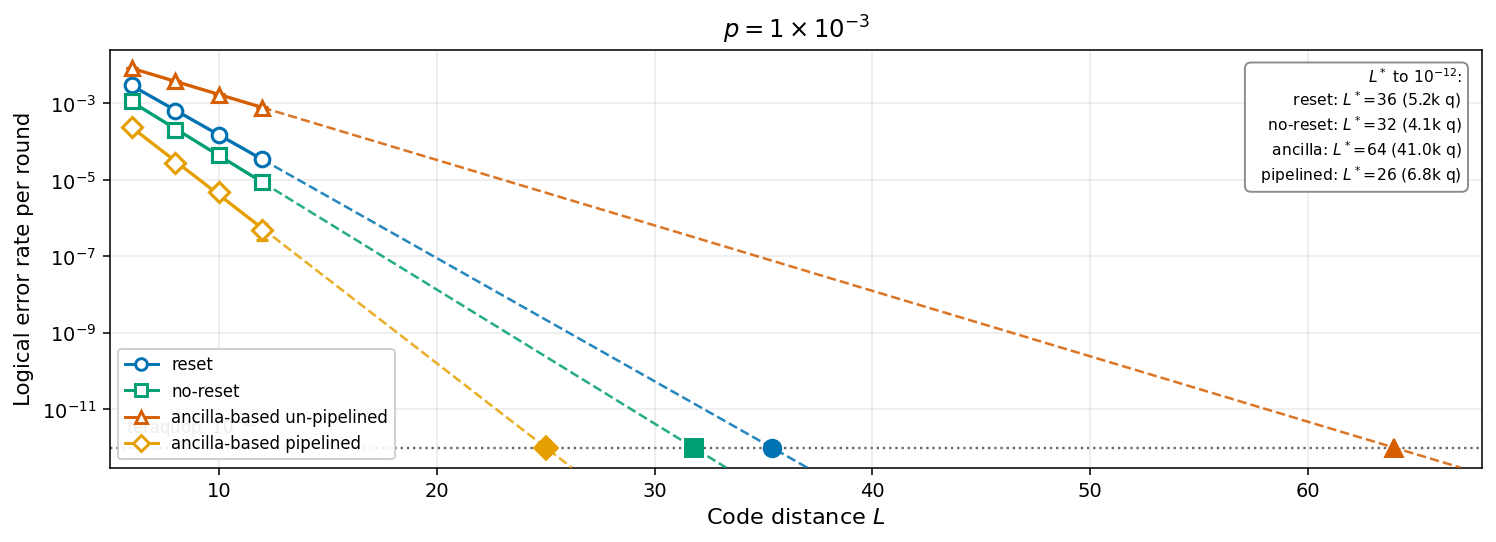}
\caption{Sub-threshold scaling $\mathrm{LER}(L)$ at $p = 1 \times 10^{-3}$ with MWPM. Markers are measured per-round LER for $L \in \{6, 8, 10, 12\}$, lines are linear fits in $\log_{10}\mathrm{LER}$ vs $L$ (slopes give $\Lambda$), extrapolated as dashed lines to the teraquop target $\mathrm{LER} = 10^{-12}$.}
\label{fig:subthreshold}
\end{figure*}

Sub-threshold scaling describes how the LER falls with code distance. Below threshold, the standard suppression ansatz is
\begin{equation} 
\mathrm{LER}(L) \approx A \, \Lambda^{-L/2}, 
\end{equation} 
where $\Lambda$ is the logical error suppression factor. Each increase of two in the code distance multiplies the LER by $1/\Lambda$. The exponent is $L/2$ because, with circuit-level distance $d = L$, the dominant logical failures come from roughly $d/2$ faults~\cite{gidney_planar_2022}. Taking $\log_{10}$, \begin{equation} \log_{10}\mathrm{LER}(L) = a - \frac{L}{2}\log_{10}\Lambda, \end{equation} so $\Lambda$ is recovered from a linear fit of $\log_{10}\mathrm{LER}$ against $L$. The scan covers $L \in \{6, 8, 10, 12\}$ for all four variants. Each $(L, p, \mathrm{variant})$ cell uses $1\times10^{7}$ shots when decoded with MWPM, and $1\times10^{5}$ shots when decoded with BP+matching.

At $p = 1 \times 10^{-3}$ (Fig.~\ref{fig:subthreshold}), the pipelined ancilla-based variant suppresses errors most strongly, with $\Lambda \approx 7.6$ -- about $1.5\times$ the next-best value -- reflecting that pipelining reduces the number of error locations during each QEC round. The un-pipelined ancilla-based variant, unsurprisingly, suppresses errors most weakly ($\Lambda \approx 2.2$) owing to its extra idling ancilla qubits. The two dynamic variants fall in between, with the no-reset suppressing more strongly ($\Lambda \approx 5.0$) than the reset ($\Lambda \approx 4.4$). Comparing the two directly, the no-reset has a higher threshold ($0.512\%$ versus $0.463\%$), a steeper sub-threshold slope, and a lower per-round LER than the reset across all simulated $(L, p)$. It is therefore the stronger of the two dynamic schemes on error suppression. Extrapolating the fits to the teraquop regime ($\mathrm{LER} = 10^{-12}$, enough to support $\sim\!10^{12}$ logical operations) gives the required code distance and qubit footprint. The pipelined variant reaches the target at the smallest distance ($L^\ast \approx 26$), but because the ancilla-based circuits need $10L^2$ physical qubits against $4L^2$ for the dynamic variants, the no-reset variant has the smallest footprint ($L^\ast = 32$, $\approx 4.1 \times 10^3$ qubits), ahead of reset ($L^\ast = 36$, $\approx 5.2 \times 10^3$), pipelined ($\approx 6.8 \times 10^3$), and the un-pipelined ancilla ($L^\ast = 64$, $\approx 4.1 \times 10^4$).

\section{Gadget mapping}
\label{app:gadget-mapping}
The two-colouring of Section~\ref{sec:two-colouring} gives an $\{A, B\}$ partition of the data graph, and the dynamic gadget chooses one class to measure in each sub-round. Naively, this is a 6-bit choice, but schedule symmetries collapse it to two independent bits -- the class measured on $XX$ sub-rounds and the class measured on $ZZ$ sub-rounds, equivalently, the global parity of the 2-colouring and the class measured on $X$-flavour rounds -- for a total of four options. I checked all four are equivalent at a range of $L$ values using $10^4$ shots per option and configuration, at $p=1.5\times 10^{-3}$ and $p=3\times 10^{-3}$; the largest pairwise spread was $2.3\sigma$, within Monte Carlo noise. The choice, therefore, does not matter for the threshold.

\section{Reduced detectors}
\label{app:detector-audit}
The reducer is purely algebraic and does not know what a plaquette is. It treats each detector as a bit-string over its measurement records and performs greedy $\mathrm{GF}(2)$ elimination, at each step keeping whichever combination has lower Hamming weight. Because it optimises for minimum weight rather than geometry, it lands on whatever is lightest, which is mostly not shaped like a plaquette. The basis has the expected weights in $\{4,8\}$, but only a small number of them are the recognisable single-plaquette detectors that match the natural Floquet-code detectors~\cite{hastinghaah2021, davydova2022}. The rest, including the weight-8 occurrences, are $XOR$ mixtures of inferences drawn from more than one plaquette. Every such mixture lies in the same detector space as the natural plaquette detectors and encodes the same error information, so the two
bases are interchangeable for decoding. The reduction is lossless by construction -- because it only takes $XOR$s of detectors, it cannot change the space they span, and the reduced reset basis has the same $\mathrm{GF}(2)$ rank as the raw flow output. The low-weight basis itself is not unique, and a different reduction ordering produces different representatives of the same space, with the retained local basis differing by only $O(1)$ detectors between orderings. The logical error rate agrees within Monte Carlo error across orderings, so the threshold is independent of this chosen basis. The count is also as expected. Each noisy Floquet period adds exactly $4L^2$ independent detectors, which I confirmed by measuring how the rank grows with $n_{\mathrm{qec}}$: one $X$-type and one $Z$-type consistency detector for each of the $2L^2$ plaquettes ($L^2$ green squares, $L^2/2$ red and $L^2/2$ blue octagons), i.e.\ $2\times 2L^2 = 4L^2$. The only detectors ever removed are the non-local ones, where dropping the topological detectors lowers the rank by exactly the number dropped, and nothing local is lost.

The no-reset variant does not need the reducer because its detectors can be read directly. Without that reset, a measured qubit keeps the state the measurement left it in. We can measure the same plaquette twice, and if nothing disturbs it in between, the two readings must agree — and we have our detector. In the period-six CSS schedule, this holds for the gap-four transitions, so one detector per plaquette and flavour is emitted as the $XOR$ of its two gap-four-separated inferences. The basis with weights in $\{4,8\}$ then falls out directly.


\section{Pipelined ancilla-based schedule}
\label{app:pipelining}
The pipelined schedule adapts the SD6 honeycomb pipelining~\cite{gidney_honeycomb_2021} to the ancilla-based 4.8.8 Floquet code. Pipelining is possible in our case because the 4.8.8 data graph, like the honeycomb code, is bipartite. The additional packing I achieve beyond what SD6 does on the honeycomb comes from the CSS schedule's lack of $YY$ measurements. The maximum pipeline depth is exactly three. Each colour appears in two sub-rounds per period (e.g.\ $rXX$ at sub-round 0 and $rZZ$ at sub-round 3), so sub-round $k$ and sub-round $k + 3$ share ancillas and a depth-4 pipeline would conflict on those ancillas. The schedule is fixed by three rules, applied modulo 8 within each cycle:
\begin{itemize}
  \item \textbf{Colour-to-TICK offset.} Red operations start at TICK 0, green at 1, blue at 2.
  \item \textbf{Phase ordering within a colour's slot.} Each colour gets a 4-TICK slot in which the four phases run sequentially: $R$, CX$_a$, CX$_b$, $M$.
  \item \textbf{Second-occurrence shift.} Sub-rounds 3-5 start 4 TICKs after the corresponding first occurrence. 
\end{itemize}

Thus, per TICK, bipartiteness ensures no data-qubit collisions, and colour-class separation ensures no ancilla-qubit collisions. The result is 8 TICKs per Floquet period in steady state, a $3\times$ depth reduction over the un-pipelined 24 TICKs. The steady-state period is two back-to-back 4-TICK colour slots, $8 = 4 \times 2$, where the factor of 4 is the in-slot sequence $R$, $CX_a$, $CX_b$, $M$, and the factor of 2 is the two sub-rounds of each colour per period, which share ancillas and so cannot be in the same slot. A single 4-TICK slot -- the naive $6\times$ reduction ($24/4$), with all six sub-rounds overlapped at pipeline depth 6 -- is blocked by this ancilla sharing. The two same-colour sub-rounds ($k$ and $k+3$) cannot overlap, capping the achievable depth at 3 (one occurrence per colour). For reference, the SD6 honeycomb achieves 6 TICKs per period on the period-3 honeycomb (i.e.\ 2 TICKs per sub-round). This depth advantage matters because under depolarising noise, idle qubits accumulate errors during noisy TICKs, so fewer TICKs per period means less idle noise per round.

To check that I applied the pipelining correctly and did not unintentionally produce a different code, I compared the instantaneous stabiliser groups (ISGs) and Pauli flows of the pipelined and un-pipelined circuits over a full Floquet period. Using Stim's \texttt{flow\_generators()}, I encoded the stabilisers and their measurement dependencies as binary symplectic vectors and verified that both circuits generated the same GF(2) row spans. The ISGs and sign-aware flows matched, showing that the pipelined circuit implements the same code and measurement structure as the original circuit, with a smaller spacetime volume.

\section{Spatial-distance verification}
\label{app:distance-verification}
I use the circuit-level definition of code distance throughout to mean the minimum number of physical errors to produce an undetectable logical error. I tried two independent ways to check that $d = L$:
\begin{enumerate}
  \item Stim's \texttt{shortest\_graphlike\_error}, on each of two independent X-cycle observables (horizontal and vertical non-contractible cycles). The result is that $d = L$ at all values of $L$ on both cycles, all four variants. 
  \item \texttt{search\_for\_undetectable\_logical\_errors} with no graphlike approximation, at various $L$ values. The result here is also that $d = L$ exactly for all four variants.
\end{enumerate}

\section{Timelike-distance verification}
\label{app:timelike-c}
The timelike distance numbers in Section~\ref{sec:results-timelike} come from a graph analysis of each circuit's DEM, as described in Appendix~B of~\cite {claes2025}. Here, I describe the procedure, the bracketing, and a few subtleties specific to the 4.8.8 case.
 
\subsubsection{Setup}
For each variant, I build the noisy circuit with $n_{\mathrm{qec}}$ Floquet periods of standard depolarising noise sandwiched between two noiseless warmup periods and two noiseless tail periods. \verb|circuit.detector_error_model(decompose_errors)| extracts the DEM (set to \verb|False| so that hyperedges are not split into pairs), which is converted into two graphs that bracket the true timelike distance $d_t$:
\begin{itemize}
  \item $g_{\mathrm{graph}}$ keeps only weight-$1$ and weight-$2$ errors as edges, discarding hyperedges. Removing connections can only lengthen the shortest path, so $g_{\mathrm{graph}}$ \emph{over}-estimates $d_t$: $d_{\mathrm{graph}} \ge d_t$.
  \item $g_{\mathrm{hyper}}$ additionally replaces each weight-$w$ hyperedge by a clique on its $w$ detectors. Now, adding connections can only shorten the shortest path, so $g_{\mathrm{hyper}}$ \emph{under}-estimates $d_t$: $d_{\mathrm{hyper}} \le d_t$.
\end{itemize}
The clique edges need not correspond to any single physical error, so $d_{\mathrm{hyper}}$ is a lower bound rather than an exact value. Together, $d_{\mathrm{hyper}} \;\le\; d_t \;\le\; d_{\mathrm{graph}}$, which is tight when the two coincide. I take \emph{early} detectors to be those in the first sub-round window of the noisy region and \emph{late} detectors those in the last, reading sub-round times from the DEM's \texttt{SHIFT\_COORDS} annotations (period index $\times 6$, plus the sub-round within the period). The bound is the shortest path in errors connecting any early detector to any late detector.
 
I scan $n_{\mathrm{qec}}$ and read the per-period rate $d_t/n_{\mathrm{qec}}$ off the closed forms. The data lie exactly on these forms for $n_{\mathrm{qec}}\ge 4$ (reset: from $n_{\mathrm{qec}}\ge 2$). The slopes are therefore exact integers (or $3/2$), and the offset constants are fixed finite-window corrections from the warmup/tail and detector-window boundaries that do not affect the rate. 

\subsubsection{Reset vs no-reset}
\begin{figure}
\centering
\includegraphics[width=\linewidth]{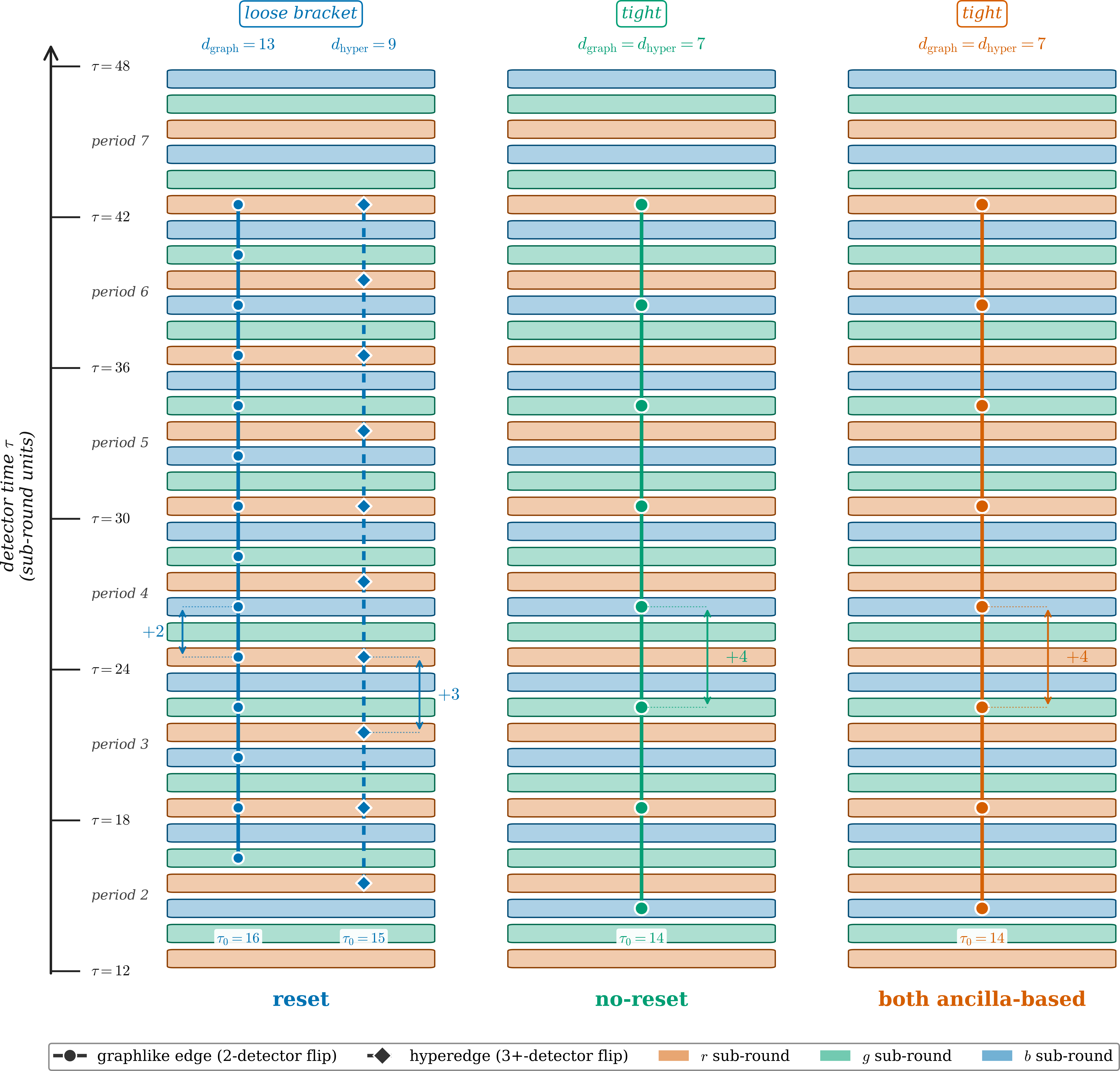}
\caption{Shortest error chain in the timelike direction, shown for an example $n_{\mathrm{qec}}=6$. The vertical axis is detector time $\tau$ in sub-round units, and the stacked $r/g/b$ tiles show the six-sub-round Floquet schedule. Graphlike (2-detector) errors are circles joined by solid lines, hyperedge ($\ge 3$-detector) errors are diamonds joined by dashed lines. The reset variant has a loose bracket, $d_{\mathrm{graph}}=13$ via $\Delta\tau=+2$ jumps and $d_{\mathrm{hyper}}=9$ via $\Delta\tau=+3$ hyperedge jumps. The no-reset and ancilla-based circuits (un-pipelined and pipelined) collapse to a tight $d_{\mathrm{graph}}=d_{\mathrm{hyper}}=7$ via $\Delta\tau=+4$ jumps, with no shorter hyperedge shortcuts. The chains start at slightly different $\tau_0\in\{14,15,16\}$ because the shortest-path search connects any early-window detector ($\tau\in[12,18)$) to any late-window one ($\tau\in[42,48)$) and picks the best-aligned phase -- these offsets are not physically meaningful and the path can change with $L$, but the repeated jump size is fixed and gives the asymptotic slope.}
\label{fig:shortest_path}
\end{figure}

The difference between the reset and no-reset timelike distances comes down to what can go wrong on the contracted qubit in a single sub-round, and how that enters the detector structure. It can be understood in the morphing-circuit language, made explicit by the shortest timelike error chain (Fig.~\ref{fig:shortest_path}).

In the reset gadget, each check uses $\mathrm{CX}\to M\to R\to\mathrm{CX}$, where the contracted qubit is measured and then reset. So two independent things can fail -- the measurement can report the wrong bit, or the reset can prepare the wrong state -- and the two flip different detectors. With both available, one can build a timelike error that alternates wrong measurement, wrong reset, wrong measurement, and so on down a single qubit. This is like the alternating string mechanism from~\cite{shaw_optimising_2026}. It advances two sub-rounds per error, so crossing a six-sub-round Floquet period costs three errors asymptotically, giving the slope-$3$ upper bound $d_{\mathrm{graph}} = 3n_{\mathrm{qec}} - 5$. The reset detector error model is not graphlike, and some single errors flip three or more detectors. Clique-expanding these hyperedges opens a faster three-sub-round step, giving the slope-$2$ lower bound $d_{\mathrm{hyper}} = 2n_{\mathrm{qec}} - 3$. The distance is therefore bracketed. These per-period slopes are those implied by the morphing-circuit contraction counting and the separated/coinciding-bound structure of~\cite{shaw_optimising_2026}, Eq.~(19) (they leave the offset constants as $O(1)$). The bounds do not coincide because $d_{\mathrm{hyper}}$ may use clique shortcuts that no single chain of hyperedge errors realises (so it can underestimate $d_t$), while $d_{\mathrm{graph}}$ ignores genuine hyperedge shortcuts (so it can overestimate). An exact value needs a stability experiment~\cite{Gidney2022stability}. As I did not implement boundaries, such a stability experiment could not be run to determine $d_t$ precisely across all variants. I leave this to future efforts. 

The no-reset gadget drops the reset, $\mathrm{CX}\to M\to\mathrm{CX}$, and with it the second failure mode. A measurement error now only flips the recorded bit, and the qubit keeps the state the measurement left it in -- exactly as in an ancilla-based check -- so there is no wrong-reset partner to alternate with, and the chain above cannot form. The cheapest timelike error is then a chain of wrong check outcomes, the same mechanism in a bare-ancilla circuit. A single wrong outcome flips two detectors four sub-rounds apart ($\tau \to \tau+4$ in Fig.~\ref{fig:shortest_path}), so the chain advances four sub-rounds per error, giving the tight slope
\begin{equation}
  \frac{d_t}{n_{\mathrm{qec}}} = \frac{6}{4} = \tfrac{3}{2},
  \qquad d_t = \big\lfloor \tfrac{3}{2} n_{\mathrm{qec}}\big\rfloor - 2,
\end{equation}
shared by the no-reset and both ancilla-based circuits with $d_{\mathrm{graph}}=d_{\mathrm{hyper}}$. This path is already graphlike, so it appears in both the graph-only and the clique-expanded detector graphs. The tight bound is not because the no-reset model has no hyperedges (it has many), but because those hyperedges span at most three sub-rounds and so never shorten the four-sub-round chain. In the language of~\cite{shaw_optimising_2026}, removing that reset eliminates the alternating error mechanism, collapsing the bracket back to the bare-ancilla value rather than improving the logical error rate at fixed distance, as it does in their biased-noise setting.
\vspace{15pt}

\section{Non-local detectors}
\label{app:nonlocal-detectors}
The Floquet schedule on a torus produces two kinds of deterministic measurement combinations. The first are \emph{local detectors}, small $XOR$s of measurement bits whose support is bounded independently of $L$. The second kind is \emph{non-local detectors}, deterministic $XOR$s supported on non-contractible cycles of the torus (e.g., at $L = 6, 8, 10, 12, 14, 16$, these have weights $72, 74, 22, 38, 18, 16$, respectively). Two of these are the global $\mathbb{Z}_2$ symmetries of the code, and the rest are loop-like detectors winding around the torus. They are not logical operators -- their value is fixed by the schedule and initialisation, independent of the encoded state -- and the matching decoder cannot use them, so I drop them from the syndrome basis. Keeping them would add high-degree nodes to the matching graph without improving error correction.

\end{document}